\documentclass[prd,letterpaper,twocolumn,superscriptaddress,preprintnumbers,nofootinbib]{revtex4-1}

\pdfoutput=1
\usepackage{graphicx}
\usepackage{amsmath,amssymb}
\usepackage{bbm}
\usepackage[T1]{fontenc}
\usepackage{float}
\usepackage{braket}
\usepackage{tikz}
\usepackage{xcolor}
\usepackage{color}
\usepackage{bbold}
\usepackage{dsfont}
\definecolor{naviBlue}{RGB}{0,0,128}
\usepackage[colorlinks  = true,
            linkcolor   = naviBlue,
            urlcolor    = naviBlue,
            citecolor   = naviBlue,
            anchorcolor = naviBlue]{hyperref}

\usepackage{amssymb}
\usepackage{slashed}
\usepackage{physics}
\usepackage[compat=1.1.0]{tikz-feynman}
\usepackage{tikz}

\newcommand{\M}{M_N}
\newcommand{\nn}{\nonumber}
\newcommand{\sE}{E}

\newcommand{\sGamma}{\Gamma}
\newcommand{\rGamma}{\tilde\Gamma}

\newcommand{\stheta}{\theta}

\newcommand{\diff}{\mathrm{d}}
\renewcommand{\vec}[1]{\boldsymbol{#1}}

\definecolor{darkgreen}{rgb}{0.0, 0.7, 0.0}

\usepackage[normalem]{ulem}

\begin{document}

\title{Probing Solar Heavy Neutrinos with Heliospheric Electrons}

\author{Marco Drewes}
\email{marco.drewes@uclouvain.be}
\affiliation{Centre for Cosmology, Particle Physics and Phenomenology, Universit\'e catholique de Louvain, Louvain-la-Neuve B-1348, Belgium}

\author{Jan Heisig}
\email{heisig@physik.rwth-aachen.de}
\affiliation{Institute for Theoretical Particle Physics and Cosmology, RWTH Aachen University, \\ 
D-52056 Aachen, Germany}

\author{Valentin Weber}
\email{valentin.weber@uclouvain.be}
\affiliation{Centre for Cosmology, Particle Physics and Phenomenology, Universit\'e catholique de Louvain, Louvain-la-Neuve B-1348, Belgium}

\begin{abstract}
We search for an excess of electrons and positrons in the interplanetary space from the decays of heavy neutrinos produced in nuclear reactions in the Sun. Using measurements of the electron spectra in the MeV range from the Ulysses and SOHO satellites, we report the strongest direct upper bound to date on the mixing between heavy neutral leptons with MeV masses and electron neutrinos, reaching $U_e^2\simeq 10^{-6}$ at $M_N=10\,$MeV\@. Our sensitivity is predominantly constrained by the uncertainties in the propagation of electrons and positrons, particularly the diffusion coefficient in the inner Solar System, as well as the uncertainties in the astrophysical background. Enhancing our understanding of either of these factors could lead to a significant improvement in sensitivity.
\end{abstract}

\maketitle

\section{Introduction}

Neutrinos are among the most enigmatic elementary particles in physics. The experimental discovery of neutrino oscillations have firmly established their nonzero masses and therewith evidence for physics beyond the Standard Model (SM).  In addition to the observed flavor mixing, neutrinos may also mix with new singlet fermions, often called sterile neutrinos. A non-vanishing mixing between the active neutrinos of flavor $\alpha$ and sterile neutrinos induces SM weak interactions of the latter with amplitudes suppressed by their mixing angles $\theta_{\alpha}$. Heavy sterile neutrinos, thus, constitute a type of heavy neutral leptons (HNLs) with  unknown mass $\M$ and interaction strength $U_\alpha^2=|\theta_\alpha|^2$ to the SM\@.
This interaction is sometimes referred to as  a \emph{neutrino portal}
~\cite{Alekhin:2015byh, Curtin:2018mvb, Beacham:2019nyx, Agrawal:2021dbo}.

In this work, we constrain the properties of HNLs with masses in the MeV range from the non-observation of an excess in the fluxes of MeV electrons and positrons in the interplanetary space of the Solar System. Such emission would be expected from the decay of long-lived HNLs that were produced in nuclear reactions in the Sun into electron-positron pairs \cite{Toussaint1981}. The range of HNL masses is limited to $1 \ {\rm MeV} \lesssim \M  \lesssim 16$ MeV, where the lower bound comes from the kinematic requirement to allow the HNLs to decay into $e^+ e^-$-pairs while the upper bound is given by the maximal energy of Solar neutrinos. The idea to search for decays of HNLs produced through nuclear reactions in the Sun has previously been considered in Refs.~\cite{Back:2003ae,Borexino:2013bot} and~\cite{Gustafson:2023hvm}, but the former analysis used the ground-based Borexino experiment while the latter considered decays into photons.\footnote{The Sun can also be used as a source in searches for other exotic long-lived particles, see e.g.~\cite{Gustafson:2023hvm,Batell:2009zp,Schuster:2009fc,Arina:2017sng,Cuoco:2019mlb}.}

HNLs in the MeV mass range are interesting candidates for addressing open questions in physics, as they can, in principle, simultaneously explain the origin of neutrino masses and the matter-antimatter asymmetry in the Universe \cite{Canetti:2012kh,Domcke:2020ety}. Upper bounds on the $U_\alpha^2$ from direct searches at accelerators or nuclear reactors are comparably weak in this mass range \cite{Britton:1992xv,Hagner:1995bn,PiENu:2015seu,Bryman:2019ssi,Bryman:2019bjg} (though some improvement is possible  \cite{PIONEER:2022yag,vanRemortel:2024wcf,Alves:2024feq,Knapen:2024fvh}). Among the various indirect constraints~\cite{Antusch:2014woa,Drewes:2015iva,Fernandez-Martinez:2016lgt,Chrzaszcz:2019inj}, the only ones that would potentially be competitive come from neutrinoless double $\beta$-decay (see e.g.~\cite{Blennow:2010th}). These can, however, largely be avoided in realistic models, in which the light neutrino masses are protected from large contributions from HNL mixing by an approximate lepton number conservation, see \cite{Asaka:2011pb,Lopez-Pavon:2012yda,Chrzaszcz:2019inj,Bolton:2019pcu,Dekens:2024hlz,deVries:2024rfh}.

Strong bounds exist from cosmological considerations \cite{Sabti:2020yrt,Boyarsky:2020dzc,Vincent:2014rja,Diamanti:2013bia,Poulin:2016anj,Domcke:2020ety}, in particular big bang nucleosynthesis (BBN). Combining these bounds with those from laboratory searches, strongly constrains the properties of HNLs that are lighter than kaons \cite{Drewes:2015iva,Drewes:2016jae,Chrzaszcz:2019inj,Domcke:2020ety,Bondarenko:2021cpc}. The resulting bounds, however, vary widely depending on model assumptions. First, they depend on the number of HNL flavors.  While the existence of HNLs that are lighter than the pion ($m_{\pi} = 139$~MeV) is essentially excluded in minimal models with only two HNL flavors~\cite{Drewes:2016jae,Bondarenko:2021cpc}, HNL lifetimes can be kept short enough to avoid stringent BBN constraints for $\M$ in the MeV range if a third flavor is added~\cite{Domcke:2020ety}. Secondly, some of the direct search bounds strongly depend on the relative mixing of HNLs with different SM flavors~\cite{Tastet:2021vwp}. Thirdly, many constraints can be avoided if the HNLs can decay into a hidden sector are possible~\cite{deGouvea:2015euy,Fischer:2019fbw,Abdullahi:2023ejc}. Finally, bounds from supernovae \cite{Mastrototaro:2019vug,Carenza:2023old} rely on explosion modeling, which introduces uncertainty in the constraints \cite{Chauhan:2023sci}, as seen with axion-like particles \cite{Bar:2019ifz}.These considerable uncertainties and dependencies on the model assumptions underscore the importance of pursuing direct HNL searches that provide complementary constraints in a controlled environment that are robust against some of the aforementioned caveats. Nuclear reactors on Earth offer one such setting \cite{vanRemortel:2024wcf}. Here, we propose a complementary approach, using the Sun as a large natural reactor and the inner Solar System as an extensive fiducial volume for detection. The current data from space-based detectors like Ulysses~\cite{Ulysses} and SOHO~\cite{SOHO} provide an independent and competitive probe of MeV-scale HNLs, the sensitivity of which is, as we find, primarily limited by the modeling of astrophysical backgrounds and could considerably improve with a better understanding of the latter.

\section{The model}

Right-handed neutrinos $\nu_R$ represent a popular extension of the SM that predicts the existence of HNLs. They are not only motivated by the observation that all other known elementary fermions have right-handed partners (needed for anomaly freedom in many gauge extensions of the SM), but also by the fact that they appear in theoretically appealing neutrino mass models that incorporate a type-I seesaw mechanism~\cite{Minkowski:1977sc, GellMann:1980vs, Mohapatra:1979ia, Yanagida:1980xy, Schechter:1980gr, Schechter:1981cv}. Moreover, they can potentially resolve several open questions in particle physics and cosmology \cite{Drewes:2013gca,Abdullahi:2022jlv}, such as the matter-antimatter asymmetry in the Universe \cite{Canetti:2012zc} through leptogenesis \cite{Fukugita:1986hr} or the dark matter puzzle \cite{Dodelson:1993je,Shi:1998km} (see  Refs.~\cite{Davidson:2008bu,Bodeker:2020ghk,Klaric:2021cpi} and \cite{Drewes:2016upu,Boyarsky:2018tvu} for reviews).

In the minimal scenario, which we consider here, the HNL's only interaction with the SM is due to their coupling to the SM weak currents \cite{Shrock:1980ct,Shrock:1981wq}, which is suppressed by the elements $|U_\alpha|\ll 1$ of the complete neutrino mixing matrix (with $\alpha = e, \mu, \tau$).\footnote{While this is usually thought of as an effective low energy description, in principle, it can be a valid description up to the Planck scale \cite{Bezrukov:2014ina}. For specific choices of the HNL parameters, known as Neutrino Minimal Standard Model \cite{Asaka:2005an,Asaka:2005pn} this permits one to simultaneously explain neutrino masses, dark matter and the origin of baryonic matter in the universe \cite{Canetti:2012kh,Ghiglieri:2020ulj}.} The HNLs are described by four-spinors $N$. For the purpose of our analysis we consider a single flavor of Majorana HNLs, $N\simeq \nu_R + \theta^T \nu_L^c + {\rm c.c.}$, which exclusively couples to electrons ($\theta_\mu = \theta_\tau = 0$). While this does not represent a fully consistent model of neutrino masses, it can approximately capture many aspects of the phenomenology of realistic models and provides well-defined benchmark~\cite{Drewes:2022akb}. For our purpose, the coupling of HNLs to weak gauge bosons ($W$ and $Z$) and Higgs bosons $h$ is captured by the phenomenological Lagrangian \cite{Atre:2009rg}:
\begin{align}\label{eq:weak WW}
 \mathcal L \supset&
- \frac{m_W}{ v} \overline N \theta_e^* \gamma^\mu e_{L} W^+_\mu 
- \frac{m_Z}{\sqrt 2 v} \overline N \theta_e^* \gamma^\mu \nu_{L e} Z_\mu \nonumber\\
  &- \frac{\M}{v\sqrt{2}}  \theta_e h \overline{\nu_L}_e N
\ + \ \text{h.c.}
\ + \ \mathcal{O}[\theta^2].
\end{align}
with $m_Z$, $m_W$ as the weak gauge boson masses and $v\simeq 174$ GeV as the Higgs field vacuum expectation value. The mass scale $\M$ is unknown; while it is traditionally associated with values near the scale of Grand Unification, neutrino oscillation data can be explained even for $\M\sim$ eV \cite{deGouvea:2005er}, and technically natural models with $\M$ below the electroweak scale exist (see~e.g.~Sec.~5 in Ref.~\cite{Agrawal:2021dbo} and references therein). In the mass range $\M \sim $ MeV that we consider here, one can integrate out the weak gauge bosons and work in Fermi theory.

\section{Heavy neutrino production and decay}

In this simple scenario the HNL production and decay are governed by only two parameters $\M$ and $U_e^2$.

\subsection{Decay rates}

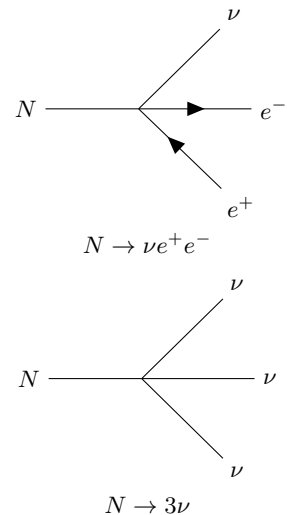
\begin{figure}[!tbp]
  \centering
  \begin{minipage}[b]{0.45\textwidth}
    \begin{tikzpicture}
      \begin{feynman}
        \vertex (a) {\(N\)};
        \vertex [right=of a] (b);
        \vertex [above right=of b] (c) {\(\nu\)};
        \vertex [below right=of b] (d) {\(e^{+}\)};
        \vertex [right=of b] (e) {\(e^{-}\)};
        \diagram* {
          (a) -- (b),
          (b) -- (c),
          (b) -- [anti fermion] (d),
          (b) -- [fermion] (e),
        };
      \end{feynman}
    \end{tikzpicture}
    \\
    \centering $N \to \nu e^{+} e^{-}$
    \vspace{6pt}
  \end{minipage}
  \hfill
  \begin{minipage}[b]{0.45\textwidth}
    \begin{tikzpicture}
      \begin{feynman}
        \vertex (a) {\(N\)};
        \vertex [right=of a] (b);
        \vertex [above right=of b] (c) {\(\nu\)};
        \vertex [below right=of b] (d) {\(\nu\)};
        \vertex [right=of b] (e) {\(\nu\)};
        \diagram* {
          (a) -- (b),
          (b) -- (c),
          (b) -- (d),
          (b) -- (e),
        };
      \end{feynman}
    \end{tikzpicture}
    \\
    \centering $N \to 3\nu$
  \end{minipage}
  \vspace{10pt}
  \caption{Relevant 3-body decays of the HNL\@.}
  \label{fig:FeynmanDiags}
\end{figure}

In the mass range of interest, Majorana HNLs can either decay fully invisibly into three active neutrinos, $N \to 3\nu$, or into a neutrino and an electron-positron pair, $N \to e^+ e^- \nu$, the latter requiring $M_N>2m_e\simeq 1.02\,$MeV\@. The Feynman diagrams of these 3-body decays are shown in Fig.~\ref{fig:FeynmanDiags}. The corresponding decay widths read:
\begin{align}
    \rGamma_{N \to 3 \nu} &= \frac{G_F^2 M_N^5}{96 \pi^3} U_e^2, \\
    \rGamma_{N \to \nu e^+ e^-} &= \frac{G_F^2 M_N^5}{192 \pi^3} U_e^2 \Lambda\!\left(\frac{m_e}{M_N}\right), \label{eq:elposdecaywidth}
\end{align}
where $\Lambda(m_e/M_N)$ is a phase-space factor that can e.g.~be found in Ref.~\cite{Gorbunov:2007ak} and we neglect the masses of active neutrinos.
Note that we label the quantities in the HNL's rest frame with a tilde.
The total decay rate at leading order is $\rGamma =\rGamma_{N \to e^+ e^- \nu} + \rGamma_{N \to 3\nu}$, 
resulting in a proper lifetime of
\begin{equation}\label{LifeTime}
 \tau \simeq 0.13 \, \text{s} \,\left(\frac{10~\text{MeV}}{M_N}\right)^5 \left(1.4 \ U_{e}^2 +  U_{\mu}^2 + U_{\tau}^2 \right)^{-1}\,.
\end{equation}
For example, the lifetime for $M_N=10\,$MeV, $U_e^2=10^{-6}$, and $U_\mu^2= U_\tau^2=0$ is roughly a day.
We do not consider loop-induced decays $N\to \nu \gamma $, which only lead to a small correction of $\tau$ \cite{Pal:1981rm,Barger:1995ty}.

\subsection{HNL flux}

Due to mixing with active neutrinos, HNLs can be produced by nuclear reactions in the Sun in the same way as Solar neutrinos are, provided the reaction is sufficiently energetic to accommodate the HNL mass. The HNL flux is therefore proportional to the flux of active neutrinos $\varphi_{\nu}$ and the mixing, $U_e^2$. At distance $R$ from the Sun the HNL flux is~\cite{Borexino:2013bot}
\begin{align}
   &\!\frac{\diff\varphi_{N}}{\diff E_N}(\sE_N, R)\; =  \,  \frac{R_\nu^2}{R^2} \mathrm{e}^{-\frac{R}{v_N} \sGamma}\int \!\diff E_\nu\,\frac{\diff\varphi_{\nu}}{\diff E_\nu}(\sE_\nu, R_\nu)\, U_e^2 \, \label{eq:phiN} \\
    &\qquad\qquad\quad\times\delta(E_\nu - E_N) \Theta(E_N-M_N) \sqrt{1 - \left(\frac{M_N}{\sE_N}\right)^2}, 
    \nn
\end{align}
where $R_\nu$ denotes the distance of neutrino flux measurement and $\sGamma = \rGamma/\gamma_N $ is the total HNL decay rate in the Solar frame. The HNL velocity and boost factor are denoted by $v_N$ and $\gamma_N$, respectively. Note that, here and in the following, the quantities without a tilde are defined in the Solar frame. The last term in Eq.~\eqref{eq:phiN} represents the phase space suppression factor due to the HNL's mass.

We consider the Solar neutrino flux as given in Ref.~\cite{Vitagliano:2019yzm}. In the energy range above the electron-positron threshold, $M_N>2m_e$, its most relevant contribution stems from ${}^8\text{B}$ decays, i.e.~the reaction $^8\text{B} \to {}^8\text{Be} + e^+ \, \nu_e$. 
\begin{figure}[H]
    \centering
    \includegraphics[width=0.9\linewidth]{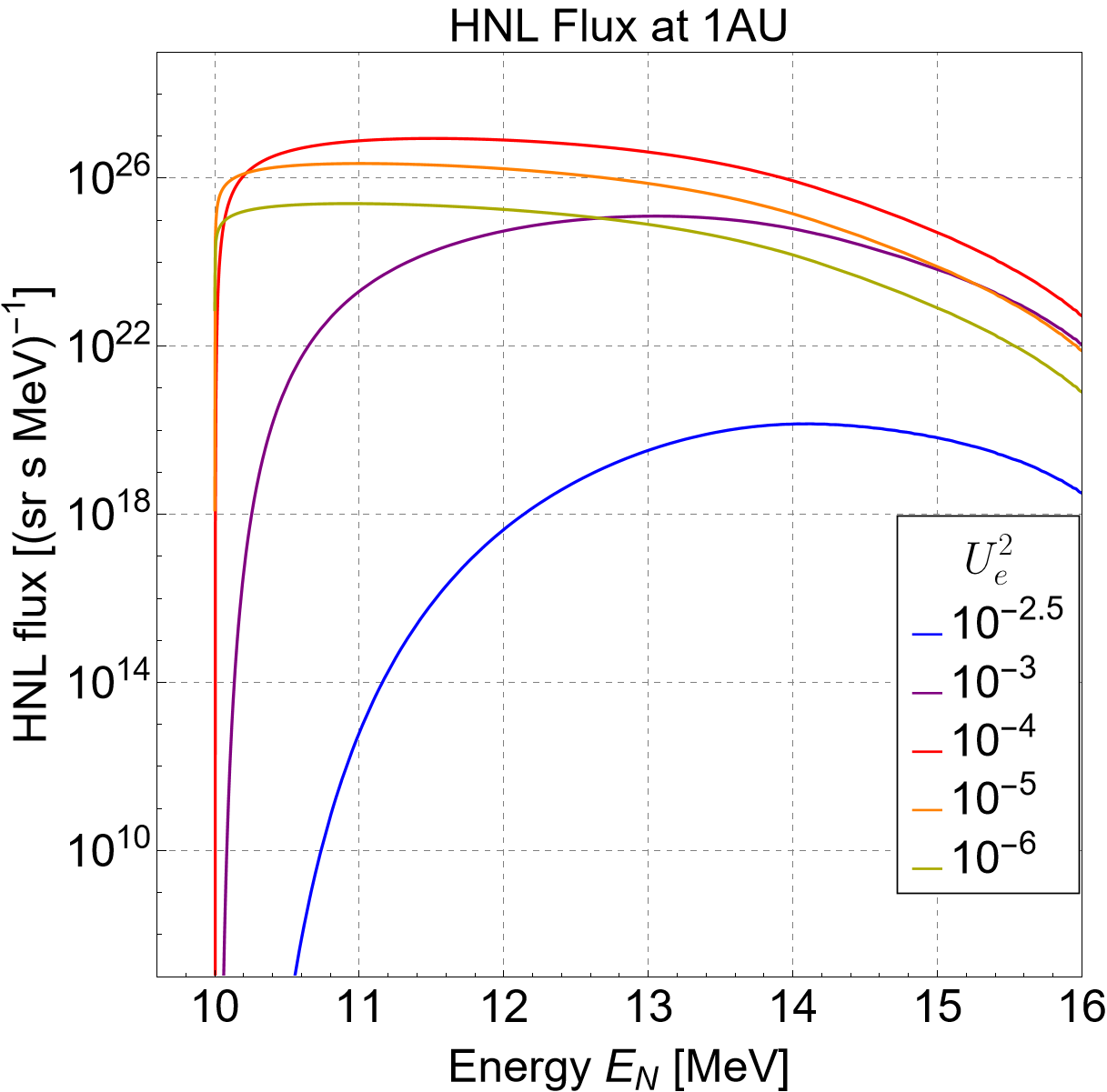}
    \caption{HNL flux $\diff\varphi_{N}/\diff E_N$ 
    at $R=1\,$AU for $M_N= 10 $ MeV and different values of $U_e^2$.}
    \label{fig:HNLFluxU2}
\end{figure}
In Fig.~\ref{fig:HNLFluxU2}, we show the fluxes for $M_N=10\,$MeV, $R=1\,$AU, and for various values of $U_e^2$. For a small mixing parameter, $U_e^2\lesssim 10^{-5}$, the $U_e$-dependence is dominantly stemming from the HNL production and hence resembles the linear scaling in $U_e^2$ from Eq.~\eqref{eq:phiN}. For larger $U_e^2$ the HNL width becomes compatible to (and eventually larger than) the inverse distance from the Sun such that the exponential suppression due to HNL decays [see Eq.~\eqref{eq:phiN}] becomes significant. Note that for smaller (larger) masses the exponential suppression sets in for higher (smaller) values of $U_e^2$.

\subsection{Positron and electron flux}
\label{sec:elecflux}

We now turn to the derivation of the observable flux of electrons and positrons from HNL decays within the heliosphere. Without loss of generality, we consider electrons (the results for positrons are identical). 

The motion of charged particles in the Solar System is heavily influenced by Solar and interplanetary magnetic fields. Directly calculating their individual trajectories within this environment is practically unfeasible due to the immense computational demands of modeling such a large ensemble of particles. Consequently, their propagation is typically described using an effective framework based on transport equations that incorporate spatial  diffusion~\cite{PARKER19659}. In the present case of MeV electrons energy losses through adiabatic cooling, resonant interactions with plasma waves, or radiation losses are small (maximally around 10\%, see~\cite{Vogt_2022} for an analysis of Jovian electrons). In the absence of such energy losses, the diffusion equation reads (see e.g.~\cite{Engelbrecht:2022gzz}):
\begin{align}
    \label{eq:diffusion}
	 \frac{\partial \psi}{\partial t} = \nabla \cdot (\vec{D} \cdot \nabla \psi) - \vec{v}_\mathrm{c} \cdot \nabla \psi + q
\end{align}
where $\psi$ is the electrons' phase space density. The first two terms on the right hand side describe diffusion, characterized by the diffusion tensor $\vec{D}$ 
and convection, characterized by the Solar wind velocity $\vec{v}_\mathrm{c}$. Note that to good approximation, $\vec{D}$ is energy-independent in the considered energy range~\cite{Potgieter:2015jxa}. The last term, $q$, denotes the source term. It is proportional to the loss in the HNL flux and the fraction of HNLs decaying into electron-positron pairs:\footnote{
We neglect effects of HNLs that are gravitationally captured within the diffusion volume ($v_N < v_{\text{esc}}$) as we estimate this contribution to be sub-leading. Although gravitationally captured HNLs decay entirely within the diffusion volume -- lifting the $U^2$-suppression for their decay probability -- this gain does not compensate for the small fraction of HNLs with $E_N < E_{\text{esc}}$. A refined analysis of this contribution is left for future work.
}
\begin{align}
    q (R,\sE_{e}) &= \int \diff\sE_N  \,\diff \cos\stheta \,\left(- \frac{\diff \varphi_{N}}{\diff R \,\diff E_N }(\sE_N,R) \right) \label{eq:differate} \\
    &\qquad\times\frac{1}{\sGamma}\frac{\diff \sGamma_{N \to \nu e^+ e^-}}{\diff\sE_{e} \diff\cos\stheta}(\sE_N,\sE_{e},\cos\stheta) \; \Theta_{\text{limits}}\,,\nn
    \\ &=\int \diff\sE_N  \,\diff \cos\stheta \, \frac{1}{v_N}\; \frac{\diff \varphi_{N}}{\diff E_N }(\sE_N,R)   \label{eq:differate2}\\
    &\qquad\times \frac{\diff \sGamma_{N \to \nu e^+ e^-}}{\diff\sE_e \diff\cos\stheta}(\sE_N,\sE_{e},\cos\stheta)\, \Theta_{\text{limits}}\,,\nn
\end{align}
where $\diff \sGamma_{N \to \nu e^+ e^-}/ (\diff\sE_{e} \,\diff\cos\stheta)$ is the double differential decay rate  in the Solar frame and $\Theta_{\text{limits}}$ contains the integration limits. The respective expression can be found in appendix~\ref{app:deffdecayrate}.

In general, the diffusion equation~\eqref{eq:diffusion} can only be solved numerically. However, as we aim for an approximate description, we can identify a set of simplifying assumptions under which an analytic solution of Eq.~\eqref{eq:diffusion} becomes possible.
First, we assume spherical symmetry effectively reducing the diffusion equation to a one-dimensional equation in the radial coordinate, $R$. Second, we assume spatial diffusion (parametrized by the diffusion coefficient $D$) to be $R$-independent and neglect convection. Third, we approximately factorize the source term into an $R$- and $E_e$-dependent part:
\begin{align}
	q(R, E_e) \simeq  \xi(E_e) \times \frac{1}{R_0 R^2} \mathrm{e}^{-R/ R_0}.
\end{align}
with 
\begin{align}
    \label{eq:R0def}
    R_0^{-1}= \frac{\int\diff E_N\,\diff R\, R^2 \,\frac{\Gamma}{v_N}\frac{\diff\varphi_{N}}{\diff E_N}(\sE_N, R) }{\int\diff E_N\diff R\, R^2 \,\frac{\diff\varphi_{N}}{\diff E_N}(\sE_N, R)}
\end{align}
and $\xi(E_e) = \big\langle q \,R_0 R^2 \mathrm{e}^{R/ R_0}\big\rangle_V$, where $\langle \,.\,\rangle_V$ denotes the diffusion volume average, weighted by the $R$-dependence of the source term, $R^{-2} \mathrm{e}^{-R/ R_0}$.
Finally, assuming a steady-state condition, $\partial \psi/\partial t=0$, we find the analytic solution:
\begin{align}
\label{eq:psianalytic}
\psi(R,E_e) =  \frac{\xi(E_e)}{D R}\, \left(1- \mathrm{e}^{-R/R_0}   -   \frac{R}{R_0}\,\mathrm{Ei}(-R/R_0)\right) \,,
\end{align}
where $\mathrm{Ei}(-R/R_0)$ is the exponential integral function. The two integration constants of the second-order differential equation have been fixed by the condition of vanishing density at $R\to\infty$ and $R_0\to \infty$, respectively.\footnote{These boundary conditions correspond to the limit $R_V\to\infty$ of the steady-state conditions for a finite diffusion volume with radius $R_V$, $\psi(R_V)=0$ and $\psi'(R_V)=-\xi \,(1-\mathrm{e}^{-R_V/ R_0}) D^{-1} R_V^{-2}$, respectively. At $R=1\,$AU, this limit provides a 10\%-level approximation to the numerical solution with $R_V= 120\,$AU, i.e.~a finite diffusion volume extending to the heliopause.} The result \eqref{eq:psianalytic} is proportional to $D^{-1}\propto (\mbox{residence time})$ as expected to (approximately) hold true for diffusion-dominated transport.

Note that a numerical solution of the spherically symmetric diffusion equation including $R$-dependent diffusion and convection can lead to sizable corrections up to around half an order of magnitude. However, similar or larger corrections are expected to arise from effects of anisotropic diffusion requiring the full three-dimensional modeling which is beyond the scope of this paper. See Sec.~\ref{sec:results} for our choice of an effective diffusion coefficient and a further discussion.

From the phase space density $\psi$, we obtain the electron flux per steradian as 
\begin{align}
	\frac{\diff \varphi_e}{\diff E_e}(R,E_e)=\frac{v_e}{4\pi} \, \psi(R,E_e)\,,
\end{align}
where $v_e$ is the velocity of the electron of energy $E_e$.

\section{Datasets and backgrounds}

\subsection{Data and experiments}

We want to use the electron (and positron) flux measurements in the interplanetary space to derive limits on an additional flux contribution from HNL decays. Due to the rapid decline of the flux per unit area, the strongest constraints can be obtained from probes that operate comparably close to the Sun.\footnote{We, for instance, checked that data from the Cosmic Ray Subsystem (CRS) installed on the Voyager 1 spacecraft \cite{Voyager1}, in spite of the extremely long exposure time, leads to considerably weaker constraints than the ones presented here.} Several space missions have been launched that are capable of measuring MeV electrons in the inner heliosphere, with the scientific goal of studying the Sun, the Solar wind, and their properties~\cite{SolarOrbiter,isis,Ulysses,SOHO,HEBER2005605,HEBER2001547,Fleth2023,Kollhoff2023}. For this analysis, we use data from Ulysses \cite{Ulysses} and the Solar and Heliospheric Observatory (SOHO) \cite{SOHO}. Specifically, we analyze measurements from the KET detector, which is part of the COSPIN module \cite{COSPIN} at Ulysses and the EPHIN sensor unit which is part of the COSTEP module at SOHO~\cite{COSTEP}. Both datasets have been taken at $R = 1$ AU and employ detectors using the Ionisation Energy Loss technique (Bethe-Bloch dE/dx curve measure) and hence cannot distinguish between electrons and positrons. To minimize the background from Solar electrons, we focus on data collected during periods of low Solar activity and without significant Solar flares. Additionally, we select periods with well-established magnetic connectivity between Jupiter and Earth, for which a background model of Jovian electrons has been published in Ref.~\cite{Vogt_2022}.

\subsection{Backgrounds}

There are three major sources of backgrounds for electrons in the energy range of interest ($\sim0.5$ to 16\,MeV):

\begin{enumerate}
    \item \textbf{Jovian Electrons}: These electrons are produced in the magnetosphere of Jupiter and constitute the dominant background in the energy range of interest during periods of low Solar activity within the inner heliosphere, particularly at $R \simeq 1 \ \mathrm{AU}$. Due to their continuous flux and relatively stable properties, Jovian electrons have been a subject of study in  space weather research. This background has been modeled, by comparing the flux observed near Jupiter during the flyby of the Pioneer 10 spacecraft to observations made by several probes near $R \simeq 1$ AU,\footnote{We note in passing that electrons and positrons coming from decaying HNLs would have affected this modeling. In our analysis, we account for this by simultaneously fitting the background model and a signal from HNL decay. An \emph{ab-initio} computation of the background could eliminate the necessity for this and greatly improve our sensitivity.}  see~\cite{Potgieter2018,Vogt2018} and references therein, 
    providing us with a reliable framework to account for their contribution in our analysis.

    \item \textbf{Galactic Electrons}: Galactic electrons originate from outside the Solar System. During their propagation through the heliosphere, they are deflected by the Solar magnetic field carried by the Solar wind. Additionally, they experience scattering and energy losses in the turbulent magnetic fields of the heliosheath, further diminishing their flux upon entering the inner Solar System. As a result, their contribution to the electron flux at energies below $\sim16 \, \mathrm{MeV}$ is minor compared to that of Jovian electrons~\cite{Potgieter2018}.

    \item \textbf{Solar Electrons}: The Sun produces substantial fluxes of electrons through processes such as Solar flares and coronal mass ejections. These events, while significant, are sporadic rather than continuous, making them unsuitable for inclusion as part of a steady-state background. For this reason, we exclude data from time periods dominated by Solar electron fluxes~\cite{Vogt_2022, Papaioannou2016}.

\end{enumerate}

Accordingly, we focus on the background from Jovian electrons. The modeling of this background is complicated by the chaotic nature of the interplanetary magnetic field (IMF) and Jupiter’s non-central location relative to Earth and the Sun, which makes the analytical determination of their trajectories challenging. As a result, numerical models are required to accurately describe their propagation. For our analysis, we will employ the model presented in Refs.~\cite{Vogt_2022} and \cite{Vogt2020}, though it contains significant uncertainties due to the inherent uncertainties of the IMF\@.

Jovian electrons propagate via two modes: parallel transport (along magnetic field lines) and perpendicular transport (diffusive transport across the heliosphere). Parallel transport, while less chaotic, requires a good magnetic connection between Earth and Jupiter (i.e. both need to be on the same Solar magnetic field line), making such conditions relatively rare. Both transport modes are characterized by the mean free path $\lambda$: $\lambda_\parallel$ for parallel transport and $\lambda_{\perp}$ for diffusive transport.

In this work, we focus on the parallel transport model, as it is better studied and allows for more robust parameter exploration. The mean free path $\lambda_\parallel$, representing the average distance an electron travels before being considerably affected by the magnetic field, is a key parameter. The analysis of Ref.~\cite{Vogt2020} suggest a value of $\lambda_\parallel = 0.15$, though, in general, this parameter can vary with the time and position of data taking. Accordingly, in our analysis, we treat $\lambda_\parallel$ as a free parameter, to be fitted to data.

Note that the background from positrons (that are not distinguished from electrons in the considered data) is expected to be negligible in comparison to the electron background~\cite{aslam2019} and hence is not considered here.

\section{Analysis and Results}

To derive constraints on the HNL model parameters we perform a $\Delta \chi^2$-analysis. Due to the uncertainties in the background modeling we vary $\lambda_\parallel$ as an \emph{a priori} unconstrained nuisance parameter which we profile over in the fit. 

\subsection{$\chi^2$-Analysis}

The total goodness of fit, $\chi^2$, for a given point in the HNL and background model parameter space, $\{M_N$,$U_e^2\}$, and $\lambda_{\parallel,d}$, respectively, reads:
\begin{align}\label{FullModel}
    &\chi^2(M_N,U_e^2;\{\lambda_{\parallel,d}\}) = \nn \\
    &\qquad\quad \sum_d\sum_{b_d} \frac{\big(\mathcal{S}_{b_d} (U_e^2,m)+\mathcal{B}_{b_d}(\lambda_{\parallel,d}) - \mathcal{D}_{b_d}\big)^2}{\sigma^2_{b_d}}
\end{align}
where the fluxes $\mathcal{S}_{b_d}$, $\mathcal{B}_{b_d}$, and $\mathcal{D}_{b_d}$ refer to the signal, background and data, respectively, in a bin $b_d$ of the dataset $d$.  The first sum runs over the considered datasets whereas the second sum runs over each bin of the two datasets. 

\begin{figure*}[t]
    \centering
    \includegraphics[width=0.42\linewidth]{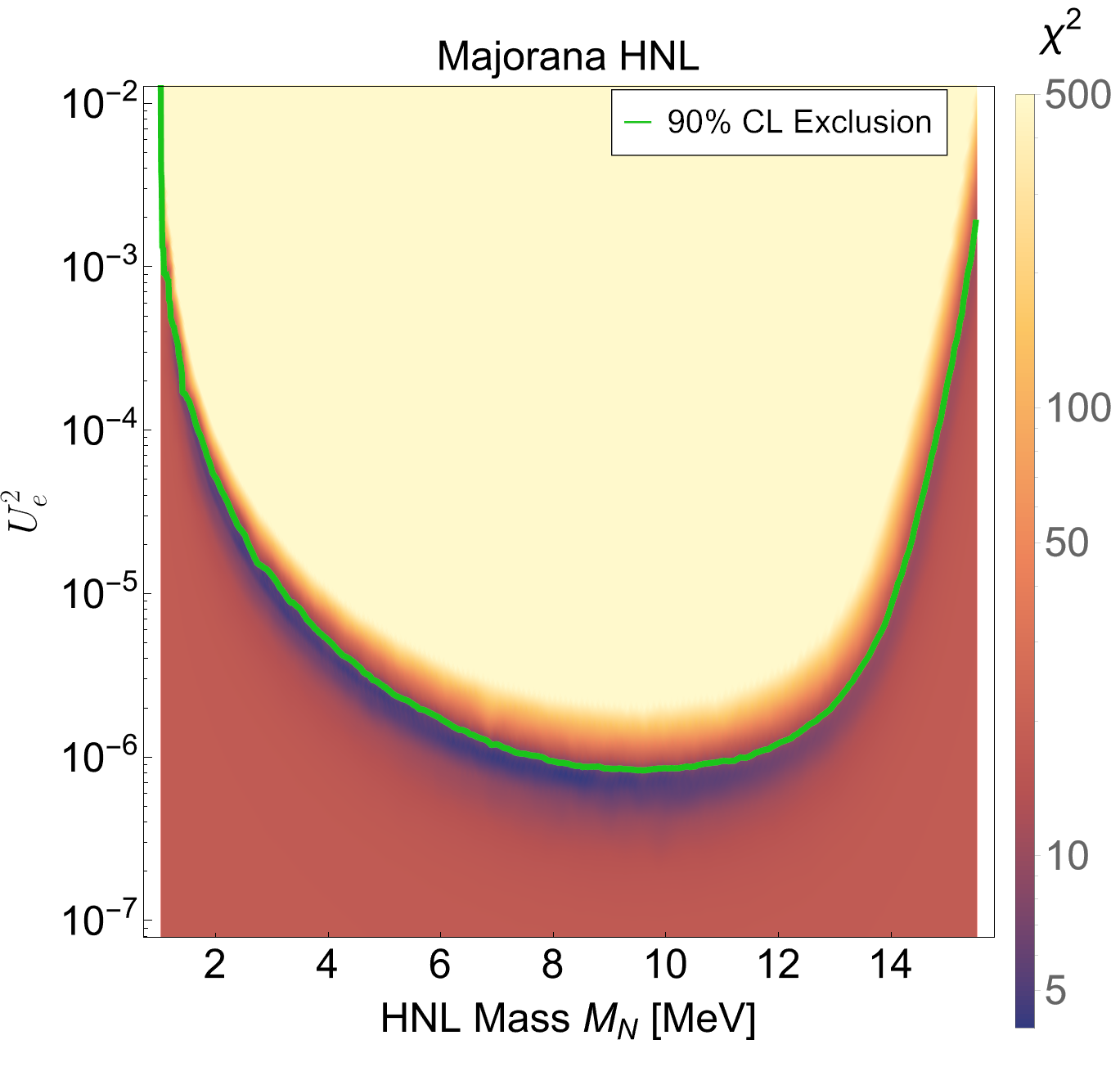}
    \hspace{4ex}
    \includegraphics[width=0.42\linewidth]{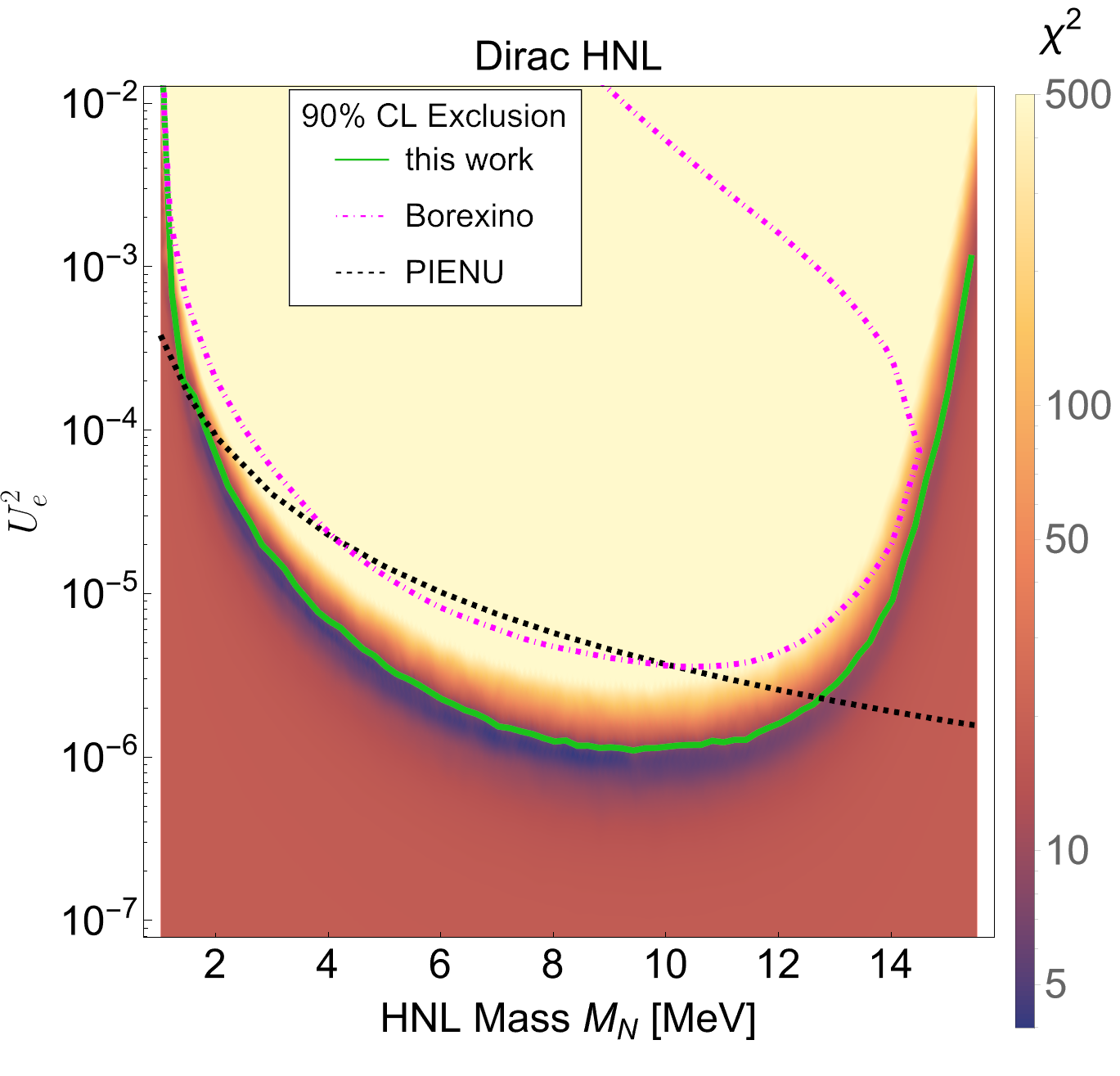}
    \caption{Constraints on the HNL parameter space in terms of the HNL mass and mixing parameter for a Majorana (left panel) and Dirac HNL (right panel). The solid green curves show the 90\%\,CL exclusion limit (the regions above the lines are excluded). The other two curves in the right panel display the upper limits from Borexino \cite{Borexino:2013bot} and PIENU \cite{Bryman:2019bjg} on $U_e^2$ for comparison. The color code displays the $\chi^2$ computed from Eq.~\eqref{FullModel}.}
    \label{fig:finalplot}
\end{figure*}

For both experiments considered, the errors $\sigma_{b_d}$ are dominated by systematics and hence proportional to the measured flux, $\sigma_{b_d}= f_d \times \mathcal{D}_{b_d}$ with $f_d$ being the relative error. The relative error is around 20\%~\cite{BerndHeber}. Specifically, we use $(20\!-\!30)\%$ for SOHO and $(10\!-\!20)\%$ for Ulysses where we consider the upper and lower boundaries as optimistic and pessimistic benchmark cases.\footnote{Note that the $\chi^2/\text{d.o.f.}$ for the background-only fit with the data of SOHO (Ulysses) is 1.9 (2.5) in the optimistic and 0.85 (0.63) in the pessimistic benchmark case.}

For each point in the model parameter space, $\{M_N,U_e^2\}$, we profile over $\lambda_{\parallel,d}$ of each dataset, denoting the values minimizing the $\chi^2$ by 
$\hat\lambda_{\parallel,d}(M_N,U_e^2)$. For the background-only hypothesis, $U_e^2=0$, the $\hat\lambda_{\parallel,d} = 0.15$ for Ulysses is consistent with the result found in Ref.~\cite{Vogt2020}. A value of $\hat\lambda_{\parallel,d} = 0.11$ is found for SOHO.

\subsection{Results}
\label{sec:results}

Following the Wilks theorem, we compute the one-sided 90\% confidence level (CL) exclusion on $U_e^2$ for a given HNL mass by considering the one-dimensional test statistic
\begin{align}
    &\Delta \chi^2 = \chi^2\big(M_N,U_e^2;\{\hat\lambda_{\parallel,d}\}\big) - \chi^2\big(M_N,\hat U_e^2;\{\hat\lambda_{\parallel,d}\}\big)
\end{align}
where $\hat U_e^2$ is the best-fit value of $U_e^2$. The result is shown in the left panel of Fig.~\ref{fig:finalplot}. The solid green line represents the 90\% CL upper bound on $U^2_e$, excluding mixing down to approximately $U^2_e \simeq 10^{-6}$ at an HNL mass of around 10\,MeV\@. 

Note that this result assumes a diffusion coefficient of $D=2\times 10^{22}\,\mathrm{cm}^2/\mathrm{s}$ in our effective one-dimensional description, Eq.~\eqref{eq:psianalytic}. This value of $D$ corresponds to parallel transport and was determined from analyses of Jovian MeV-electrons during periods of magnetic connectivity; it is consistent with the mean free path $\lambda_\parallel \sim (0.1\!-\!0.15)$ used in our background modeling~\cite{Vogt_2022,2024ApJ...961...57S} (see also \cite{Lang_2024} for a recent analysis of parallel transport from different source regions). A more comprehensive modeling approach would incorporate both parallel and perpendicular transport. Since the mean free path for perpendicular transport is significantly smaller than the parallel one (by up to two orders of magnitude)~\cite{FERRANDO1997905,Potgieter:2015jxa}, the chosen value likely leads to an underestimation of the flux. However, given additional corrections due to an $R$-dependence of $D$ (which tend to lower the predicted flux at 1\,AU but is subject to large uncertainties~\cite{2022ApJ...929....8E}) and convection effects (see Sec.~\ref{sec:elecflux}) we expect the resulting limit to be a realistic estimate. A more accurate description requires full three-dimensional modeling as well as accounting for various uncertainties in the propagation model, which is beyond the scope of this work. Note, however, that in the relevant region of small $U_e^2$, the limit on $U_e^2$ scales with the square root of the flux, keeping the impact of the discussed inaccuracies at a modest level.

\begin{figure*}[t!]  
    \centering
    \includegraphics[width=0.4\linewidth]{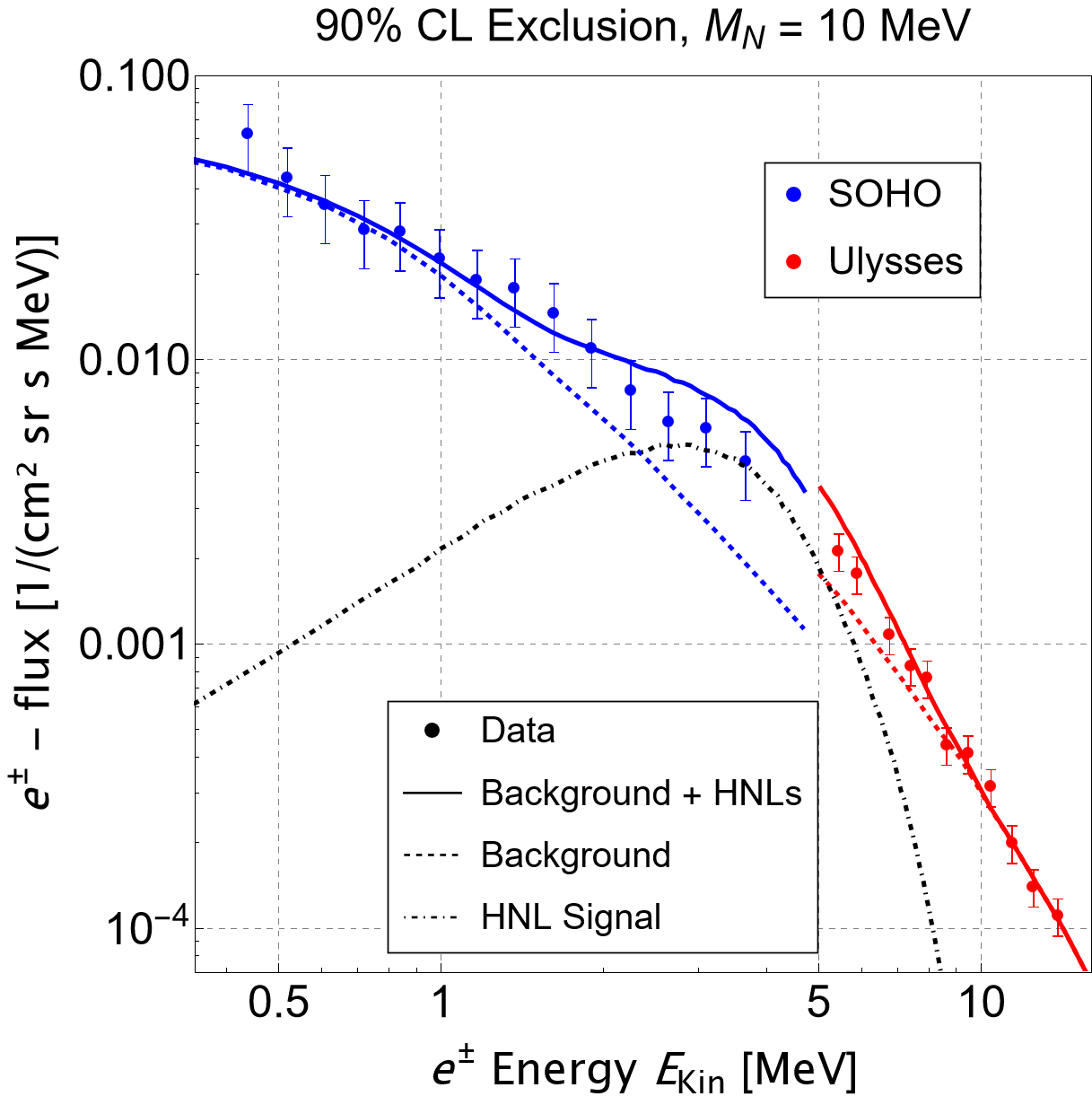}
    \hspace{4ex}
    \includegraphics[width=0.4\linewidth]{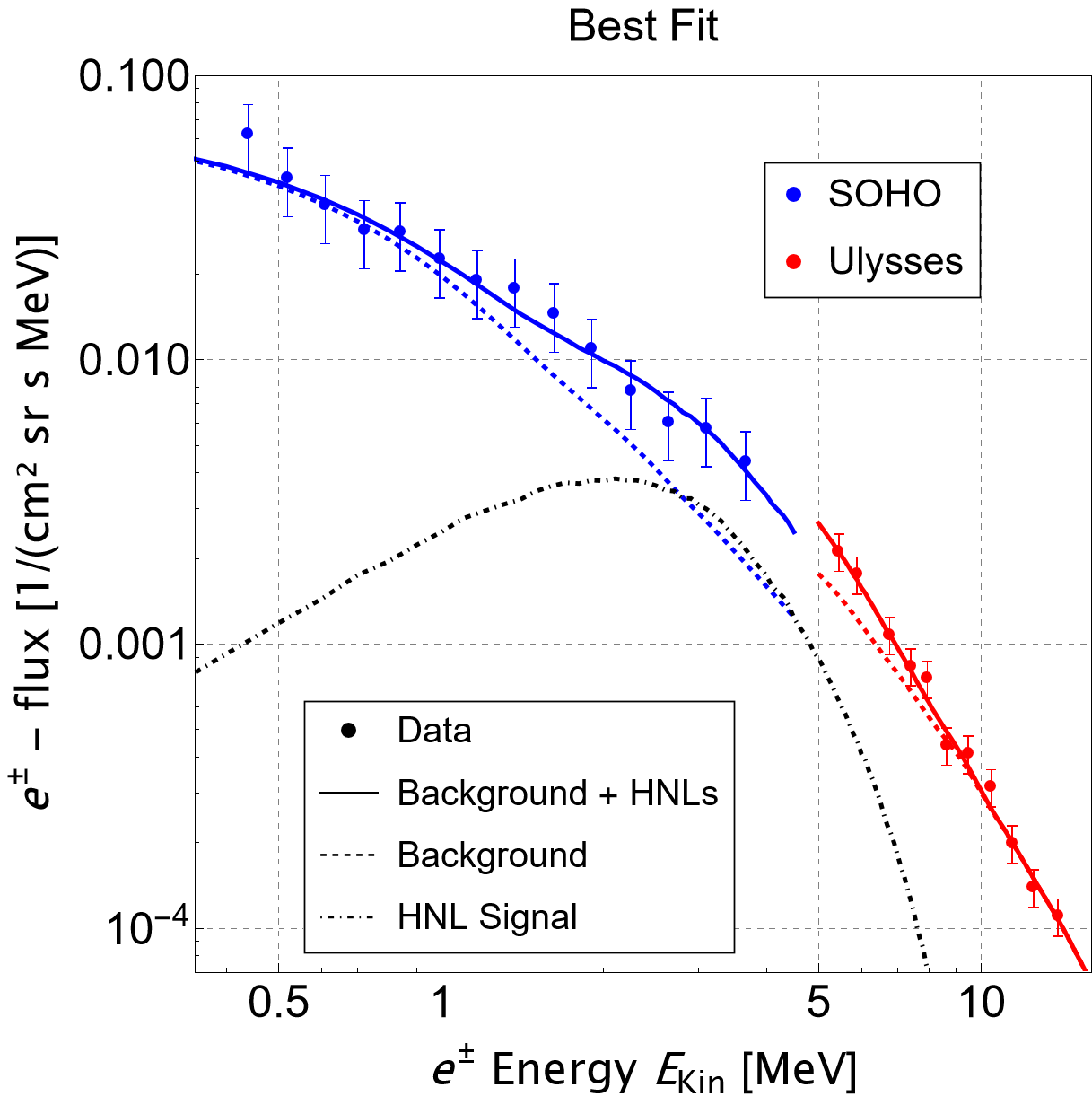}
    \caption{Flux of electrons and positrons measured by SOHO (blue data points) and Ulysses (red data points) at $R\simeq1\,$AU and the respective flux prediction from HNLs (black dot-dashed line) and fitted background model (short-dashed lines) as well as their sum (solid lines). We perform a separate background-fit for each experiment and show the respective curves in their relevant energy range only. The left and right panels show the spectra for the 90\% CL exclusion limit at $M_N=10\,$MeV (corresponding to $U^2_e\simeq 8.5\times10^{-7}$) 
    and the best-fit point ($M_N\simeq 8.2\,$MeV, $U^2_e\simeq 7.9\times10^{-7}$), respectively. 
    }   \label{fig:spectrum}
\end{figure*}

Note also that, conservatively, we consider the pessimistic scenario of experimental uncertainties. However, we find an almost identical exclusion limit for the optimistic choice. The respective spectra for the 90\% CL exclusion at $M_N=10\,$MeV are shown in the left panel of Fig.~\ref{fig:spectrum}. The two datasets provide similar contributions to the $\chi^2$.

Fig.~\ref{fig:finalplot} also displays the total $\chi^2$ as color code. Note that when interpreted as a two-dimensional $\chi^2$-distribution, it reveals a slight preference of the fit for a small signal contribution at the level of around $3\sigma$. The respective spectra are shown in the right panel of Fig.~\ref{fig:spectrum}. While this indicates tentative evidence for the existence of HNLs with $\M \simeq 8.2 \,$MeV if taken at face value, the considerable astrophysical uncertainties make an incomplete background modeling a more likely explanation.

Finally, in the right panel of Fig.~\ref{fig:finalplot}, we display the respective constraints interpreted in the case of a Dirac HNL, allowing us to draw a direct comparison to the existing results from Ref.~\cite{Borexino:2013bot}, which were obtained by searching for the very same process that we consider, but using the ground-based Borexino detector.\footnote{We were able to approximately reproduce the result from \cite{Borexino:2013bot} by computing the number of expected HNL decays from the flux \eqref{eq:phiN} within a volume of  $315\,\text{m}^3$ at $R = 1$ AU.} The analysis is identical with the difference that the total decay rate of a Dirac HNL is by a factor 2 smaller.\footnote{Note that we have neglected potential differences in the angular distribution \cite{Balantekin:2018ukw} in the differential rate that could affect the integral in Eq.~\eqref{eq:differate}. However, as our analysis does not exploit angular distributions we expect this effect to be small.} 
Our limits are stronger than the one from Ref.~\cite{Borexino:2013bot} by a factor of around 3 in most of the considered mass range. Furthermore, our analysis excludes parameter space not covered by Borexino toward large $M_N$ and small to intermediate $U_e^2$, where the HNL lifetimes are too short to provide a large signal of HNL decays inside the Borexino detector.

The striking similarity of the curves for small $U_e^2$ can partially be explained by the facts that the upper and lower ends of the sensitivity region in $M_N$ are dictated by the kinematics of nuclear reactions in the Sun and HNL decays, respectively, and that the overall shapes of the curves are related to the spectrum of Solar neutrinos $\diff\varphi_{\nu} / \diff E_\nu$. However, the fact that the maximal reach of the two exclusion lines only differ by a factor $\sim 3$ is sheer coincidence.\footnote{The effective fiducial volume of our search is over $10^8$ times larger than that of the Borexino detector. This tremendous advantage, however, appears to be largely canceled by the significantly larger background, leading to surprisingly similar sensitivities.} We note in passing that a suitable re-interpretation of the results reported in Ref.~\cite{Borexino:2013bot} is likely to lead to an upper bound on $U_e^2$ for Majorana HNLs that differs from the Dirac case only by a factor $\sim\sqrt{2}$. Hence, the relative strength of the exclusion bounds reported there and in our results 
for Majorana HNLs would roughly be the same as in the Dirac case.

Additionally, we compare our results to those reported in Ref.~\cite{Bryman:2019bjg} based on data from the PIENU experiment at TRIUMF, which were also obtained for Dirac HNLs. For intermediate masses, the respective limit is very similar to the one of Borexino, which again is coincidental, as this experiment did not only use an entirely different detector, but also a different source (pion decays at an accelerator). 

Finally, we note that loop-induced HNL decays into photons offer another avenue for direct searches, as explored in~\cite{Gustafson:2023hvm}. However, the sensitivity of this approach is considerably weaker and does not compete with any of the aforementioned results within the considered mass range.

Hence, our conservative  analysis thus provides the strongest direct limits in the HNL mass range between 2 and 12\,MeV\@.

\section{Discussion and Conclusions}

We performed the first search for signatures from HNLs produced in nuclear reactions in the Sun that decay into $e^+ e^-$-pairs using space-born detectors, probing a range of HNL masses $1 \ {\rm MeV} \lesssim M_N \lesssim 16$ MeV\@. More specifically, we used data from the Ulysses and SOHO satellites to search for an excess of electrons and positrons in the MeV range that would be expected from the decay $N \to e^- e^+ \nu$. Our upper exclusion limit on $U_e^2$ reaches down to  $U_e^2 \simeq 10^{-6}$ and is displayed in Fig.~\ref{fig:finalplot}.
It represents the currently strongest direct upper bound on $U_e^2$ in the HNL mass range $(2\!-\!12)\,$MeV, outperforming limits from Borexino and PIENU\@. The limit assumes that HNLs exclusively mix with the first SM generation and can be relaxed if this assumption is lifted. However, the same applies to the results obtained by Borexino, so that the relative strength of the bounds would be approximately the same for $U_\mu^2, U_\tau^2\neq 0$. Bounds from supernova observation can, in principle, be much stronger than the exclusion curves reported here, but suffer from large uncertainties in the modeling of the explosion, while our approach relies on the well-established modeling of nuclear reactions in the Sun. In contrast to cosmological constraints in the same mass range, our analysis does not rely on any assumptions about the early universe.

In the future, satellite-based searches for HNLs from the Sun can be improved in several ways. First, one can include a larger dataset. For example, the Solar Orbiter mission provides the Energetic Particle Detector (EPD) \cite{SolarOrbiter}, and the Parker Solar Probe (PSP) offers the Integrated Science Investigation of the Sun (IS$\odot$IS) module \cite{isis}, both of which can detect electrons in the relevant energy range. Secondly, an \emph{ab initio} modeling of the electron backgrounds would help to increase the sensitivity of existing satellites to HNLs\@. This includes the modeling of the propagation of electrons and positrons which constitutes an important ingredient of the signal prediction but are currently subject to considerable uncertainties. Finally, utilizing detectors that can distinguish positrons from electrons in the considered energy range could significantly reduce the backgrounds, which are up to three orders of magnitude smaller for positrons \cite{aslam2019}. However, space probes that are able to distinguish positrons from electrons, such as PAMELA \cite{PAMELA:2013vxg} and AMS-02 \cite{PhysRevLett.110.141102,PhysRevLett.121.051102}, have only been launched to near-Earth orbits strongly affected by the geomagnetic cut-off and hence only sensitive to larger energies.

\begin{acknowledgments}
We thank Eugene Engelbrecht, Yannis Georis, Jan Gieseler, Bernard Heber, Patrick K\"uhl, Quentin Loriau, Sarah Mechbal, Philipp Mertsch,  and Marius Potgieter for valuable discussions. 
We further thank Alessandro Cuoco, Eugene Engelbrecht and Georg Raffelt for feedback on our manuscript. 
J.H.~acknowledges support from the Alexander von Humboldt Foundation through the Feodor Lynen Research Fellowship for Experienced Researchers and the Feodor Lynen Return Fellowship during the early stage of this work.
Computational resources have been provided by the supercomputing facilities of the Universit\'e catholique de Louvain (CISM/UCL) and the Consortium des \'Equipements de Calcul Intensif en F\'ed\'eration Wallonie Bruxelles (C\'ECI) funded by the Fond de la Recherche Scientifique de Belgique (F.R.S.-FNRS) under convention 2.5020.11 and by the Walloon Region.

\end{acknowledgments}

\begin{appendix}

\section{Differential decay rate and integration limits}
\label{app:deffdecayrate}

In this appendix, we provide details on the quantities used in Eq.~\eqref{eq:differate} that are not displayed in the main text.
First, for the boosted differential decay rate of HNLs with momentum $p_N$ into electrons with momentum $p_e$ in the Solar frame, we obtain:
\begin{widetext}
\begin{align}\label{GammaNSolarFrame}
\frac{\diff \sGamma_{N \to \nu e^+ e^-}}{\diff E_e \, \diff\cos\theta} = & \frac{G_F^2 U_e^2 |\bold{p}_e| \left(M_N^2 - 2 (p_e^\mu p_{N\mu}) \right)^2}{96 \pi^3 E_N
\left(m_e^2 + M_N^2 - 2 (p_e^\mu p_{N\mu}) \right)^3} \times \Bigg[ 
    -2 m_e^2 M_N^2 \big(3 m_e^2 + M_N^2 (1 - 8 (s_\text{w}^2 + 2 s_\text{w}^4)) \big) \\
& + (p_e^\mu p_{N\mu}) \Big( 
        12 m_e^4 (1 + 2 s_\text{w}^2 + 4 s_\text{w}^4) + 
        9 M_N^4 (1 + 4 s_\text{w}^2 + 8 s_\text{w}^4) + 
        m_e^2 M_N^2 (25 + 28 s_\text{w}^2 + 56 s_\text{w}^4)\nonumber \\
& + 2 (p_e^\mu p_{N\mu}) \big( 
            -17 M_N^2 (1 + 4 s_\text{w}^2 + 8 s_\text{w}^4) - 
            6 m_e^2 (3 + 8 s_\text{w}^2 + 16 s_\text{w}^4) + 
            16 (1 + 4 s_\text{w}^2 + 8 s_\text{w}^4) (p_e^\mu p_{N\mu})
        \big)
    \Big)
\Bigg],\nonumber
\end{align}
\end{widetext}
where \( p_e^\mu p_{N\mu} = E_N E_e - |\bold{p}_N| |\bold{p}_e| \cos\theta \), and \( s_\text{w} = \sin(\theta_\text{w}) \), with \( \theta_\text{w} \) being the weak mixing angle.
The expression of Eq.~\eqref{GammaNSolarFrame} is consistent with results reported for $\tilde{\Gamma}$ in the literature \cite{Gorbunov:2007ak,Bondarenko:2018ptm,Ballett:2019bgd,Coloma:2020lgy}.

Secondly, we provide explicit expressions for the integration limits represented by $\Theta_{\text{limits}}$ in Eq.~\eqref{eq:differate}.
They are given by
\begin{align}
&E_{\text{min}} < E_N < E_{\text{max}}\,, \\
&\cos\theta_{\text{min}} < \cos\theta < 1\,,
\end{align}
where,

\begin{align}
&E_{\text{min}} = 
\begin{cases} 
    M_N & \text{if } \sE_e < \frac{M_N}{2}, \\
    \frac{M_N^2 \sE_e - M_N \sqrt{ (M_N^2 -4 m_e^2)( \sE_e^2-m_e^2)}}{2 m_e^2}
    & \text{otherwise},
\end{cases}
\end{align}

\begin{align}
&E_{\text{max}} = 
\frac{M_N^2 \sE_e + M_N \sqrt{ (M_N^2 -4 m_e^2)( \sE_e^2-m_e^2)}}{2 m_e^2},
\\
&\cos\theta_{\text{min}} = 
\max\left(
    -1,\,
    \frac{2E_N \sE_e - M_N^2}{2\sqrt{(E_N^2 - M_N^2)(\sE_e^2 - m_e^2)}}
\right).
\end{align}
We perform the integration over $\cos\stheta$ first.

\end{appendix}

\bibliographystyle{bjstyle}
\raggedright\bibliography{bibliography}

\end{document}